\documentclass[aps,prl,preprint,superscriptaddress,floatfix,12pt]{revtex4}

\usepackage{graphicx}

\begin{document}


\title{Compliant substrate epitaxy: Au on MoS$_2$}


\author{Yuzhi Zhou}
\affiliation
{Department of Materials Science and Engineering, University of California at Berkeley, Berkeley, California 94720, USA}
\affiliation
{Materials Sciences Division, Lawrence Berkeley National Laboratory, Berkeley, California 94720, USA}

\author{Daisuke Kiriya}
\affiliation{Electrical Engineering and Computer Sciences, University of California, Berkeley, California 94720, United States}
\affiliation{Materials Sciences Division, Lawrence Berkeley National Laboratory, Berkeley, California 94720, USA}

\author{E. E. Haller}
\affiliation
{Department of Materials Science and Engineering, University of California at Berkeley, Berkeley, California 94720, USA}
\affiliation
{Materials Sciences Division, Lawrence Berkeley National Laboratory, Berkeley, California 94720, USA}

\author{Joel W. Ager III}
\affiliation
{Department of Materials Science and Engineering, University of California at Berkeley, Berkeley, California 94720, USA}
\affiliation{Materials Sciences Division, Lawrence Berkeley National Laboratory, Berkeley, California 94720, USA}

\author{Ali Javey}
\affiliation{Electrical Engineering and Computer Sciences, University of California, Berkeley, California 94720, United States}
\affiliation{Materials Sciences Division, Lawrence Berkeley National Laboratory, Berkeley, California 94720, USA}

\author{D. C. Chrzan}
\affiliation
{Department of Materials Science and Engineering, University of California at Berkeley, Berkeley, California 94720, USA}
\affiliation
{Materials Sciences Division, Lawrence Berkeley National Laboratory, Berkeley, California 94720, USA}
\email[Send correspondence to:]{dcchrzan@berkeley.edu}

\date{\today}

\begin{abstract}
The epitaxial growth of \{111\} oriented Au on MoS$_2$ is well documented despite the large lattice mismatch ($\approx 8$\% biaxial strain), and the fact that a Au \{001\} orientation results in much less elastic strain. An analysis based on density functional  and linear elasticity theories reveals that the \{111\} orientation is stabilized by a combination of favorable surface and interfacial contributions to the energy, and the compliance of the first layer of the MoS$_2$.
\end{abstract}


\maketitle


The electronic and optical properties of transition metal dichalcogenides show much promise for technological applications \cite{radisavljevic2011single}.  Incorporating this material within devices will require either the growth of the dichalcogenides on other substrates, or growth of other materials on a dichalcogenide substrate.  In this respect, the growth of Au on MoS$_2$ can be viewed as a prototypical system.

The growth of Au on MoS$_2$ was studied in the mid to late 1960's using early {\em in situ} and {\em ex situ}  transmission electron microscopy  \cite{AuMoS21964,AuMoS21967,AuMoS21969,Pashley64,Jacobs66}.  Au was deposited on MoS$_2$ using evaporation, and was discovered to grow predominantly with a plate-like geometry in a \{111\} orientation.  The $\langle110\rangle$ directions of the Au platelets were nearly aligned along the $\langle 11\bar{2}0 \rangle$ direction of the substrates.  Similar orientations were observed for Ag nuclei \cite{Jacobs66}.

\begin{figure}
\includegraphics{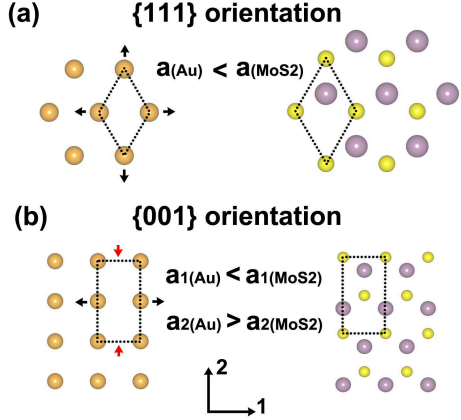}
\caption{\label{fig:fig1} (color online) (a) The strain for Au (left) and MoS$_{2}$ (right) in \{111\} orientation. The Molybdenum atoms are represented by purple circles, the Sulfur atoms by yellow circles and the Gold atoms by orange circles. a(Au) and a(MoS$_{2}$) are the lattice constants in Au (111) plane and MoS$_{2}$ respectively. The arrows indicate the biaxial strain in this case. (b) The strain in \{100\} orientation. The subscripts "1" and "2" are two perpendicular directions in the plane. a$_{1}$ and a$_{2}$ are the lattice constants in corresponding directions. In this case, the sign and amount of strain in the two directions is different.}
\end{figure}

The growth and evolution of the Au films were modeled using theory available at the time \cite{Jesser67,AuMoS21967}.  This theory, however, was rooted in an phenomenological understanding of the interfacial properties.  Moreover, the substrate was treated as a typical bulk, and no accounting for the influence of the van der Waals (VDW) bonding within the substrate layers was attempted. Further, the Au clusters were approximated as spherical caps, as the TEM images did not allow for measurement of island thicknesses.  Finally, only the \{111\} orientation was considered in any detail.

\begin{figure}
\includegraphics{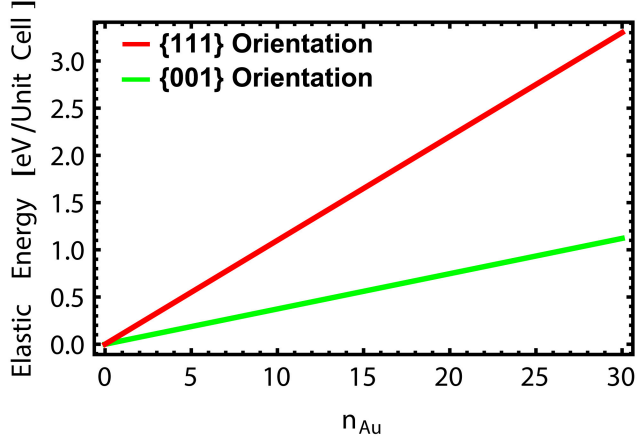}
\caption{\label{fig:fig2}(color online) The strain energy per unit cell for Au deposition on MoS$_2$ in two orientations computed assuming that the MoS$_2$ is unrelaxed, and that surface and interfacial energies make negligible contributions to the energy.  Note that the strain energy associated with the \{111\} orientation of Au exceeds the strain energy of the \{001\} oriented growth by a substantial amount.}
\end{figure}

In the most simple model of epitaxy, the substrate is assumed infinite, and as a consequence, it does not relax during the growth process.  For Au on MoS$_2$, two possible orientations of the growing film are shown in Fig.~\ref{fig:fig1}. For the \{111\} orientation, the biaxial strain required in the film is approximately 8\%.  In contrast, the \{001\} oriented film the strains are approximately -6\% and 8\% in the directions shown.  Based on these strains, one would expect that the \{001\} orientation would be much more favorable, and consequently would be predominantly the experimentally observed orientation.  Figure~\ref{fig:fig2} compares the elastic energies of the two films (neglecting surface and interfacial stresses and energies) as a function of the number of layers of Au grown assuming the substrate is rigid.  Clearly, the elastic energy of the \{111\} orientation is much larger than that of the \{001\} orientation.  This observation raises a  fundamental questions regarding the growth: Why is the predominant orientation \{111\} and not \{001\}?

In the following paragraphs a model to explain the experimentally observed film orientation is developed.  The model assumes that due to the weak VDW bonding between layers of MoS$_2$, the surface layer is able to relax nearly independently of the remaining bulk layers.  The compliance of the substrate, when coupled with the surface and interface energies (including strain energies) results in a lower formation energy for the \{111\} orientation as compared with the \{001\} orientation, in agreement with experiment.

The model is built on two types of calculations.  First, a continuum linear elastic model for the epitaxial growth of Au on MoS$_2$ including the relaxation of the substrate and the influence of surface/interface energies and strains is developed. The parameters for this model are then determined using density functional theory (DFT) based electronic structure total energy methods.

The continuum linear elastic model is developed by proposing a ``synthesis path'' and then computing the energy contributions along this path.  (Note that this path is not necessarily experimentally accessible.  It simply facilitates computation.)
The initial step in forming an epitaxial layer can be taken as the extraction of a thin slab of the growing material from a bulk crystal.  This extraction creates two surfaces, and the energy of the slab is increased by the surface energies (that reflect any strain in the as produced surfaces).  The next step is to strain the film to its final strain in its epitaxially bonded state.  This contributes the strain energy of the bulk plus any contribution to the strain energy from the surfaces.  The third step is to separate the first substrate layer from rest of the substrate to create a freestanding MoS$_2$ layer.  This adds a layer separation energy to the system.  In the fourth step, this free-standing layer is then strained introducing the strain energy of the single layer.  In the fifth step, the epitaxial film is welded to the free standing layer.  This has the consequence that one of the strained Au surfaces is replaced with a strained Au-MoS$_2$ interface.  In the final step, the film/free-standing layer is readhered to the substrate, returning the layer separation energy to the thermal bath, but introducing the energy required to slip the top MoS$_2$ layer relative to the remaining layers. Taken in total, the sum of the changes in energy for both the substrate and the epitaxial film as compared with their bulk counterparts, $\Delta E$ can be written:
\begin{equation}
\Delta E = E_{Au,sur} + E_{film} + E_{sub}
 + E_{Au/S,int} + E_{slip}
\label{eqn:elastic}
\end{equation}
where $E_{Au,sur}$ is the energy of the strained Au surface, $E_{film}$ is the energy of the strained film neglecting surface contributions, $E_{sub}$ is the strain energy of the first substrate layer, $E_{Au/S,int}$ is the interfacial and strain energy of the Au-MoS$_2$ substrate interface, and $E_{slip}$ is the slip energy between the first layer of the substrate and the remaining substrate.

The first four contributions to $\Delta E$ can be expressed analytically using linear elasticity theory:
\begin{eqnarray}
& E_{film} & =  \frac{1}{2} C_{f,ijkl} \epsilon_{f,ij} \epsilon_{f,kl} V_{f}\\
& E_{sub} & =  \frac{1}{2} C_{sub,ijkl} \epsilon_{sub,ij} \epsilon_{sub,kl} A_{sub}\\
& E_{Au,sur} & =  \gamma_{s} + f_{s,ij} \epsilon_{s,ij} + \frac{1}{2} C_{s,ijkl} \epsilon_{s,ij} \epsilon_{s,kl} A_{s}\\
& E_{Au/S,in} & =  \gamma_{I} + f_{I,ij} \epsilon_{I,ij} + \frac{1}{2} C_{I,ijkl} \epsilon_{I,ij} \epsilon_{I,kl} A_{I}
\end{eqnarray}
\noindent
The $C_{f,ijkl}$, $C_{sub,ijkl}$ are the elastic constants for the film and the top layer substrate. The strain tensors are indexed similarly. We approximate the surface/interface stress energies up to the second order of the strain tensor: $\gamma_s$ and $\gamma_{I}$ are the unstrained surface and interfacial energies, respectively; $f_{s,ij}$ and $f_{I,ij}$ are the linear surface stress terms while $C_{s,ijkl}$ and $C_{I,ijkl}$ are the quadratic terms or the effective elastic constants of the surface/interface. The top layer of the substrate is treated as a 2D material. (We assume that the VDW interaction within the substrate is sufficient to insure the substrate remains flat during the epitaxial growth). Define $V_{f}$ to be the equilibrium volume the film would have if it were part of a bulk Au crystal. $A_{sub}$, $A_{s}$ and $A_{I}$ are the reference unit cell areas for the substrate, interface and surface strains respectively. $A_{s}$ and $A_{I}$ are taken equal to $A_{f}$, the area covered by the Au film with its bulk lattice parameter. $A_{sub}$ is taken to be the equilibrium area of the monolayer MoS$_{2}$ unit cell.

The slip energy arises from displacing the top substrate layer relative to layers below. An approximation for this energy is made by investigating a similar slip in a bilayer MoS$_{2}$ system. Two MoS$_{2}$ layers are placed relative to each other in the same way as the two adjacent layers in the bulk MoS$_{2}$. The slip system is then created by straining one layer by 5\% biaxially while keeping the other one fixed. (5\% is about the typical amount of strain in the Au-MoS$_{2}$ epitaxy system.) The interlayer distance in both cases is fixed at 6.25 \AA \ , the value obtained from DFT calculations of the pristine bilayer MoS$_{2}$ system. The slip energy consists of only the change in the interlayer VDW energy since the change of the VDW energy within the strained layer is captured in the elastic energy computation. The VDW interaction energy is directly calculated following Grimme's D2 method \cite{Grimme2006}. The difference of the interlayer VDW energies between the slipped system and original system gives an estimate for the slip energy. The computed slip energy in this case is less than 2 meV per unit cell and will make a negligible contribution to $\Delta E$. (This assertion is borne out by the calculations below. For more details of the slip energy calculation please refer to the Supplemental Material.) Therefore, for simplicity, this term is neglected in the model.

The value of $\Delta E$ at equilibrium is then determined by minimizing the right hand side of Eqn. (\ref{eqn:elastic}) with respect to the strains in the film and the first substrate layer subject to the constraint that the first substrate layer and the film are lattice matched across their interface.  (A more detailed description is provided in the Supplemental Material.)

The parameters that enter the theory are determined either by fitting the results of DFT based electronic structure total energy calculations to the continuum theory or by direct DFT calculations. DFT calculations are performed using the plane-wave code VASP \cite{Kresse93}. The exchange and correlation energy are described by generalized gradient approximation proposed by Perdew, Burke, and Ernzerhof \cite{Perdew96}. Electron-ion interactions are treated with projector augmented wave potentials \cite{Perdew99}. All calculations are performed using a plane-wave basis with a 350 eV energy cutoff. The precision tag is set to ``accurate." The convergence criterion for self-consistent field loop is $1 \times 10^{-8}$ eV. A 20 \AA \ vacuum slab is added along the direction normal to the growth plane to separate the system from its periodic image.

\begin{figure}
 \includegraphics{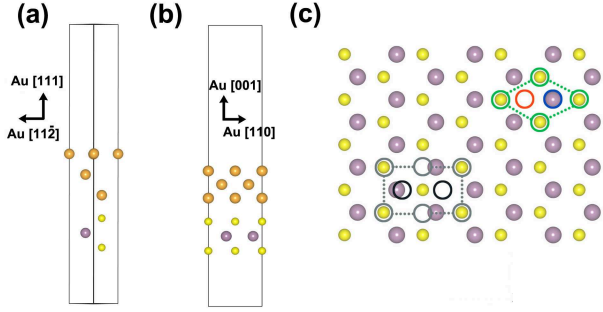}
  \caption{\label{fig:fig3} (color online) Side view of (a) \{111\} epitaxy unit cell with 3 layers of Au and (b) \{001\} epitaxy unit cell with 3 layers of Au. The vacuum layer is used to prevent the interaction of the system with its periodic image. (c) Top view of two types of the epitaxy for Au grown on MoS$_{2}$. The unit cell of \{111\} epitaxy is defined by the green dashed line . The unit cell of \{001\} epitaxy is defined by the grey dashed line.}
\end{figure}

Two epitaxial configurations are considered, as shown in Fig.~\ref{fig:fig3}. In  \{111\} epitaxy (Fig.~\ref{fig:fig3}(a) and top right of \ref{fig:fig3}(c)), Au (111) plane is lattice matched onto the MoS$_{2}$. In Fig.~\ref{fig:fig3}(c), the green circles are the gold atoms in the first layer that lie exactly on top of the Sulfur atoms. The green dashed lines define the unit cell. The gold atoms in the second and third layer are projected onto the growth plane and are represented by the blue and red circles, respectively. The stacking of gold layers follows the ``ABCABC..." sequence to be consistent with the Face Center Cubic (FCC) structure, with the ``B" sites residing on top of the Mo atoms. (The ``ACBACB" sequence is slightly higher in energy (about 0.01 eV/unit cell) than the ``ABC" configuration. Thus we take the ``ABC" sequence for \{111\} epitaxy, though both structures have been realized experimentally \cite{Jacobs66}.) In \{001\} epitaxy, the Au (001) plane is lattice matched onto the MoS$_{2}$ (Fig.~\ref{fig:fig3}(b) and bottom left of \ref{fig:fig3}(c)). In the top view, the gold atoms in the first layer are represented by grey circles. The grey dashed line defines the unit cell. The black circles are the projections of the gold atoms in the second layer. The stacking sequence follows ``A$^{\prime}$B$^{\prime}$A$^{\prime}$B$^{\prime}$..." sequence. We constructed models for both configurations with the number of Au layers vary from 3 to 30. A 14$\times$14$\times$1 Monkhorst-Pack k-point grid is used to sample the Brillioun zone for \{111\} epitaxy. In \{001\} epitaxy, a 13$\times$8$\times$1 Monkhorst-Pack k-point grid is used. These structures are relaxed until the maximum Hellmann-Feynman force on any atom is below 0.01 eV/\AA.

The DFT calculated lattice structures and elastic constants of bulk Au and monolayer MoS$_{2}$ are in good agreement with previous studies (details in the Supplementary Materials). The calculated elastic constants are used for the parameters $C_{f,ijkl}$ and $C_{sub,ijkl}$. The calculated lattice structures are used to determine $V_{f}$, $A_{sub}$, $A_{s}$ and $A_{I}$.

Within the density functional theory approach, the change in energy associated with our synthesis path can be written (neglecting the slip energy):
\begin{equation}
\Delta E = E_{tot} - n_{Au} E_{b,Au} - E_{b,MoS_{2}},
\label{eqn:DFT}
\end{equation}
with $E_{tot}$ the total energy of the lattice matched slab including one MoS$_2$ layer and $n_{Au}$ layers of Au (in each unit cell). $E_{b,Au}$ is the energy per layer of bulk unstrained Au and $E_{b,MoS_{2}}$ is the energy of the monolayer MoS$_{2}$. We fit the results of DFT calculations to both Eqn. (\ref{eqn:elastic}) and the equilibrium strains (with more weight on the energy fit side) to determine the parameters remaining in that expression. (These turn out to be linear combinations of the surface and interfacial energy terms. See the Supplemental Material for more details about fitting processes.)

\begin{figure}
\includegraphics{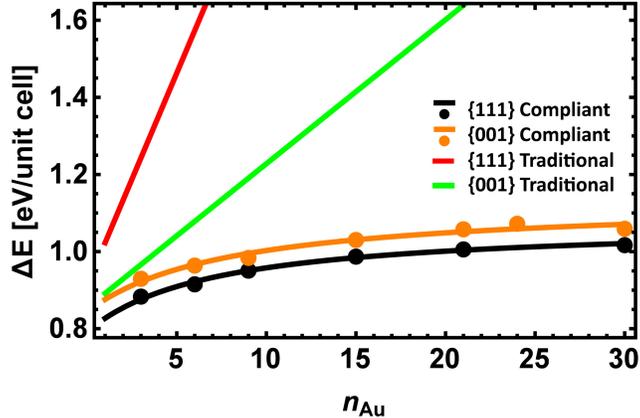}
\caption{\label{fig:fig4} (color online) A comparison between the elastic energy of compliant substrate epitaxy and that of traditional epitaxy that assumes a rigid substrate. Note that surface and interfacial energies and strain energies are included in the lines labelled traditional. DFT results are shown as dots, and the lines passing through them are the results of the continuum theory fitted to the DFT data.  Inclusion of surface/interface effects alone does not stabilize the \{111\} orientation.  This stabilization requires a compliant substrate.}
\end{figure}

Figure \ref{fig:fig4} compares the $\Delta E$'s for \{111\} and \{001\} oriented epitaxial growth. The results of DFT calculations are displayed, as is their fit to Eqn. (\ref{eqn:elastic}). It is apparent that the elasticity theory based model does an excellent job of describing the DFT results, and can be used to extrapolate behavior to thicker films. In addition, Fig.~ \ref{fig:fig4} displays the predictions assuming a rigid substrate (traditional epitaxy), and the interfacial energies and stresses computed within the model.

Examination of the Fig.~\ref{fig:fig4} reveals two important points. First and foremost, within the compliant substrate epitaxial model, for all considered numbers of Au layers, \{111\} epitaxy is energetically favored over \{001\} oriented epitaxy.  However, the energy difference is not too large, and one might expect to see both orientations, with the \{111\} orientation favored. Thus the compliant substrate epitaxy model is able to explain the experimental observations of \{111\} oriented epitaxy of Au on MoS$_2$.

Second, the variation of $\Delta E$ with thickness is sublinear in the case of compliant substrate epitaxy.  This sublinear behavior originates in the fact that the strain energy per layer is decreasing as the film thickness increases.  As the film gets thicker, it becomes elastically more stiff, and the first layer of the substrate is forced to deform to a greater extent.  Eventually, when the film is infinitely thick, only the first substrate layer deforms, and the elastic energy saturates at a constant.  To our knowledge, this reduction in strain energy of the film with increasing film thickness is a unique feature of compliant substrate epitaxy.

These observations suggest the following understanding of the preferred orientation of the films.  First, the relaxation of the substrate is not expected to change the sign of the strain energy difference between \{111\} and \{001\} oriented films. Based on strain energy alone, one would still expect \{001\} oriented films, even for a compliant substrate. This implies that the sum of the \{111\} Au surface and Au/MoS$_2$ interfacial energies is less than the sum of the \{001\} Au surface and Au/MoS$_2$ interfacial energies in the compliant substrate epitaxy, a fact consistent with our fitted parameters (See Supplemental Material). So in this instance, the properties of the interfaces and the surfaces dictate the orientation of the growing film.

This observation is not, in itself surprising.  When the films are nucleated, they are very thin, and the surfaces and interfaces can make a significant contribution to $\Delta E$.  However, the persistence of the orientation preference with increasing film thickness is remarkable.  Typically, one expects that the strain energy difference would become the dominant term in $\Delta E$, and the favored orientation would change to \{001\} for thicker films.  This is where the compliance of the substrate becomes important.  Since the substrate is compliant, the strain energy of the Au film no longer scales with the volume of the Au film.  Instead, it monotonically decreases with increasing film thickness, enabling the interfacial and surface contributions to the energy to dictate the stability of the two orientations for all film thicknesses.

The model presented here is the not first proposed that exploits a compliant substrate.  As early as 1991, Lo suggested that the quality of some epitaxial films could be improved by employing a compliant substrate \cite{Lo}.  Such a substrate would enable relaxation of the film at the expense of the substrate, but had the potential to increase the quality of the epitaxial film. Lo suggested that such substrates could be produced using standard lithographic methods.  Later, Jesser {\em et al.} proposed that a compliant substrate might be developed by introducing a subsurface twist boundary \cite{Jesser}. Here, it is noted that layered materials with VDW bonding between them form naturally compliant substrates for epitaxial growth \cite{Alaskar}.  Moreover, the VDW bonding, while enabling lateral slip of the substrate, will resist buckling of the film, and thereby help to improve its quality.  Though we have examined Au on MoS$_2$ in detail, the idea is quite general, and should apply to other systems as well.

Note that this version of compliant substrate epitaxy is different from the so-called van der Waals epitaxy \cite{Koma199272}. In the compliant substrate epitaxy model, lattice registry is maintained across the interface, and a degree of covalent bonding between the substrate and the film is allowed.  In contrast, in the van der Waals epitaxy case, one does not expect lattice registry between the film and the substrate.

We note that compliant substrate epitaxy may have interesting implications for strain engineering and processing of thin films. Consider the strain field near a small Au island in the early stages of the film growth.  The MoS$_2$ under the island will be strain in compression.  This will naturally be accommodated by a tensile strain surrounding the island.  Therefore, it might be possible to engineer the positions of the Au nuclei so as to induce a desired strain pattern into the first layer of the MoS$_2$ substrate.  Moreover, it might be possible to use the epitaxial binding as a means of freeing and manipulating the top layer of the MoS$_2$, just as was demonstrated for C nanotubes \cite{CNTTransfer}. Finally, the fact that under conditions of compliant substrate epitaxy, surface and interfacial energies can dictate the orientations of thick films raises the possibility of using surface chemistry and/or surfactants to engineer the relative stability of different film orientations within the same materials system.

In conclusion, the epitaxial growth of Au on MoS$_2$ is studied using a combination of elasticity and density functional theories.  It is shown that the compliance of the substrate in conjunction with the relative surface and interfacial energies and strain energies stabilize the \{111\} growth orientation despite the large lattice mismatch.  This system is thus an example of compliant substrate epitaxy.

\begin{acknowledgments}
This work was supported by the Director, Office of Science, Office of Basic Energy Sciences, Division of Materials Sciences and Engineering, of the U.S. Department of Energy under Contract No. DE-AC02-05CH11231.
\end{acknowledgments}

\bibliography{Zhouetal}

\end{document}



\title{Compliant substrate epitaxy: Au on MoS$_2$ \\Supplemental Materials}


\author{Yuzhi Zhou}
\affiliation
{Department of Materials Science and Engineering, University of California at Berkeley, Berkeley, California 94720, USA}
\affiliation
{Materials Sciences Division, Lawrence Berkeley National Laboratory, Berkeley, California 94720, USA}

\author{Daisuke Kiriya}
\affiliation{Electrical Engineering and Computer Sciences, University of California, Berkeley, California 94720, United States}
\affiliation{Materials Sciences Division, Lawrence Berkeley National Laboratory, Berkeley, California 94720, USA}

\author{E. E. Haller}
\affiliation
{Department of Materials Science and Engineering, University of California at Berkeley, Berkeley, California 94720, USA}
\affiliation
{Materials Sciences Division, Lawrence Berkeley National Laboratory, Berkeley, California 94720, USA}

\author{Joel W. Ager III}
\affiliation
{Department of Materials Science and Engineering, University of California at Berkeley, Berkeley, California 94720, USA}
\affiliation{Materials Sciences Division, Lawrence Berkeley National Laboratory, Berkeley, California 94720, USA}

\author{Ali Javey}
\affiliation{Electrical Engineering and Computer Sciences, University of California, Berkeley, California 94720, United States}
\affiliation{Materials Sciences Division, Lawrence Berkeley National Laboratory, Berkeley, California 94720, USA}

\author{D. C. Chrzan}
\affiliation
{Department of Materials Science and Engineering, University of California at Berkeley, Berkeley, California 94720, USA}
\affiliation
{Materials Sciences Division, Lawrence Berkeley National Laboratory, Berkeley, California 94720, USA}
\email[send correspondence to:]{dcchrzan@berkeley.edu}


\date{\today}


\maketitle

\subsection{1. The lattice structures and elastic constants of Au and MoS$_{2}$ from DFT calculations.}
The lattice structures and elastic constants of bulk Au and monolayer MoS$_{2}$ computed using density functional theory are presented in this section. The parameters for the DFT calculations are same as those reported in the main text except where noted. For the bulk Au primitive unit cell, a symmetrized 12$\times$12$\times$12 Monkhorst-Pack k-point grid is used to sample the Brillioun zone. For the primitive cell of monolayer MoS$_{2}$, a symmetrized 14$\times$14$\times$1 Monkhorst-Pack k-point grid is used to sample the Brillioun zone.

\begin{figure}[h]
 \includegraphics{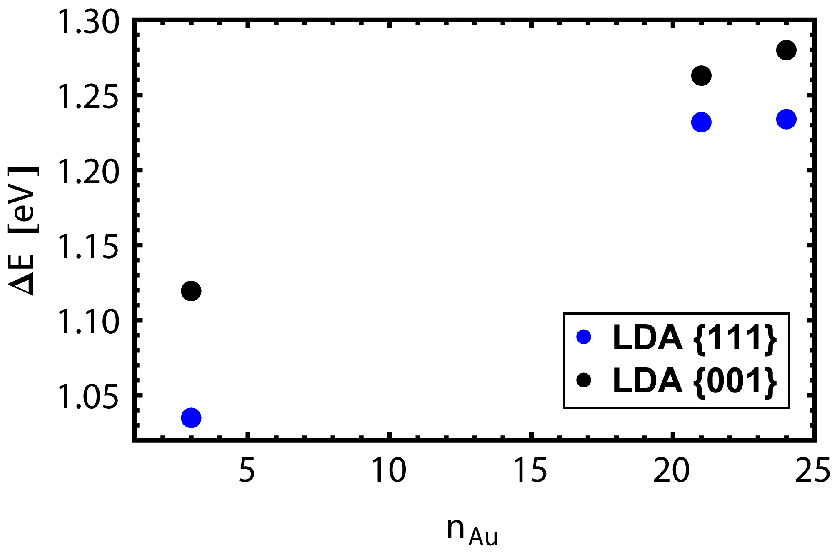}%
  \caption{\label{fig:fig1} (color online) $\Delta E$ calculated by LDA for the epitaxy systems with 3, 21 and 24 layers of Au. The \{111\} epitaxy systems are lower in energy than the \{001\} epitaxy systems.}
\end{figure}

The computed Au lattice constant 4.168 \AA, and is in reasonable agreement with experiment result of 4.062 \AA \ and in good agreement with previous DFT result of 4.154 \AA \ \cite{AuLatticeConstant}. The in-plane lattice constant of MoS$_{2}$ from our DFT calculation is 3.190 \AA \ and the thickness of the monolayer (the separation distance between the top and bottom Sulfur layers) is 3.130 \AA. These are also in good agreement with the experimental values of 3.122 \AA and 3.116 \AA, respectively \ \cite{MoS2Lattice}. The computed elastic constants for these two materials are summarized in Table~\ref{tab:table1}. Our calculated results for gold elastic constants (from GGA PBE) are softer than the experimentally measured values \cite{AuElastic}. We also calculated the Au's elastic constants using PAW LDA potential (parameterized by Perdew and Zunger) \cite{LDA}. The LDA results overestimate the elastic constants of Au, as shown in Table~\ref{tab:table1}. We further double checked our epitaxy system calculations (with 3, 21 and 24 layers of Au) using LDA. In the LDA calculations, the \{111\} epitaxy systems are still more stable than the \{001\} epitaxy systems, as shown in Fig.~\ref{fig:fig1}, which is consistent with the GGA results. We, therefore, present GGA results in the main text.

In the Ref~\cite{MoS2Elastic}, a discussion of the elastic constants of MoS$_{2}$ (both measured and predicted) is given. The C$_{12}$ measured experimentally is negative, which contrasts with the positive value computed here and with the DFT-D2 calculations done by Peelaers and van de Walle. Peelaers and van de Walle argued that the C$_{12}$ measurement is indirect, and might be complicated by the lubricating properties of MoS$_{2}$. In order to directly compare our results with those in Ref~\cite{MoS2Elastic}, we follow their convention of calculating the elastic constants for 2D material, in which the thickness of monolayer MoS$_{2}$ is chosen to be the half length of the unit cell vector perpendicular to the basal plane in bulk MoS$_{2}$ (contains 2 MoS$_{2}$ layers). Our calculated elastic constants are in good agreement with Peelaers and van de Walle's results. The calculated structures and elastic constants are used for the parameters of our continuum elastic model, {\em i.e.} the equilibrium volume of the film, the equilibrium area of the substrate/surface/interface and the elastic constants of the film and the top layer substrate.
\begin{table}
\caption{\label{tab:table1}%
The elastic constants of bulk Au and MoS$_{2}$.
}
\begin{ruledtabular}
\begin{tabular}{ccccc}
\textrm{ \hspace{20 mm} } & \textrm{Elastic Constants} & \textrm{GGA [GPa]\footnote{The GGA (PBE) results are used in the calculations described in the main text.}}& \textrm{LDA [GPa]}& \textrm{Experiment [GPa]\footnote{The previous reports of Au elastic constants can be found in Ref~\cite{AuElastic}. The reported MoS$_{2}$ elastic constants is from Ref~\cite{MoS2Elastic}. The DFT calculated elastic constnats from this reference is in parenthesis.}}\\
\colrule
\multirow{3}{*}{\textrm{Au}}  & C$_{11}$ & 154 & 219 & 192\\
                              & C$_{12}$ & 136 & 183 & 163\\
                              & C$_{44}$ & 31 & 41 & 39\\ \hline
\multirow{2}{*}{\textrm{MoS$_{2}$}} & C$_{11}$ & 211 & 237 & 238 (238)\\
                                    & C$_{12}$ & 53 & 56 & -54 (64)\footnote{Ref~\cite{MoS2Elastic} argued that the negative C$_{12}$ from the experiment is not measured directly and might come from lubricating properties of MoS$_{2}$.}\\
\end{tabular}
\end{ruledtabular}
\end{table}

\subsection{2. The slip energy estimation in MoS$_{2}$ substrate.}
\begin{figure}[h]
 \includegraphics{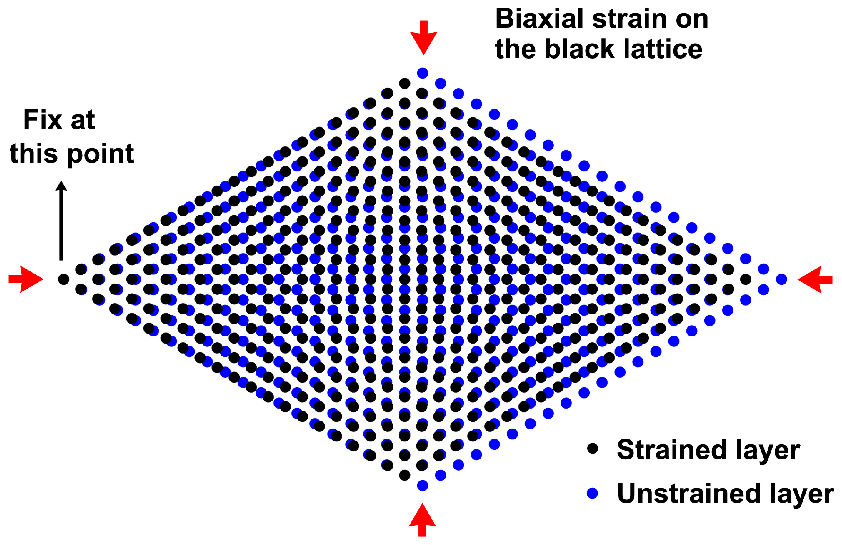}%
  \caption{\label{fig:fig2} (color online) The two layer MoS$_{2}$ system (top view). Only the lattice points are shown. The black points indicate the first layer who has 5 \% of biaxial strain. The blue points indicate the second layer which is without the strain. The lattice points of the two layers also coincide at the corners of the strained layer.}
\end{figure}

In this section, an esimate of the first layer slip energy for the MoS$_{2}$ substrate is presented. The system we use is a bilayer MoS$_{2}$ system. A top view of this system is in Fig.~\ref{fig:fig2} with only the lattice points shown. The two layers coincide at the origin point in the top view (left corner). Each layer contains 21$\times$21 unit cells. The Mo and S atoms are added to the lattice points in the same way as that in the two adjacent layers in the bulk MoS$_{2}$. The interlayer distance is 6.25 \AA, as obtained from DFT with van der Waals corrections of the pristine bilayer MoS$_{2}$ system. To create a slip similar to the compliant substrate epitaxy system, the first layer MoS$_{2}$ is strained biaxially by 5\% while the second layer is kept fixed. This is the typical amount strain in the Au-MoS$_{2}$ epitaxy system. This strain results in a structure wherein the lattice parameters of the slipped layer and the unslipped layer are commensurate (the lattice points coincide in \emph{xy} plane at the corners of the strained layer). When the strain is applied, the Mo and S atoms are also displaced with the lattice points and no internal relaxation of the unit cell is allowed.

We follows Grimme's D2 method \cite{Grimme2006} to include the van der Waals (VDW) interaction energy and calculate the interlayer VDW interaction using the formula:
\begin{equation}
E_{corr} = - \sum_{i}^{N_{at1}} \sum_{j}^{N_{at2}} \frac{C_{6ij}}{r^{6}_{ij}} f_{dmp}(r_{ij})
\end{equation}
Where the index i goes over the atoms in the first layer and j goes over the atoms in the second layer. $r_{ij}$ is the distance between atom i and atom j. $C_{6ij}$ is a coefficient which depends on the types of atom i and atom j:
\begin{equation}
C_{6ij} = \sqrt{C_{6i} C_{6j}}
\end{equation}
The parameter $C_{6i}$ and $C_{6j}$ are tabulated for each element and are insensitive to the particular chemical situation. The values of $C_{6i}$ for the elements in the first five row of the periodic table are provided in Ref~\cite{Grimme2006}. $f_{dmp}(r_{ij})$ is the damping function whose expression is:
\begin{equation}
f_{dmp}(r_{ij}) = \frac{s_{6}}{1+e^{-d(r_{ij}/(R_{0ij})-1)}}
\end{equation}
$s_{6}$ is a global scaling parameter which depends on DFT functionals used. We choose the optimized value for PBE functional ($S_{6} = 0.75$) since the structure of MoS$_{2}$ is determined by DFT calculations with PBE functional. The coefficient $R_{0ij}$ also depends on the types of atom i and atom j:
\begin{equation}
R_{0ij} =  R_{0i} +  R_{0j}
\end{equation}
The values of $R_{0i}$ for the elements in the first five row of the periodic table are also provided in Ref~\cite{Grimme2006}.

To estimate the slip energy, we compute the interlayer VDW energy of the original and slipped systems described above. The energy difference between these two values provides an estimate of the slip energy. Only the interlayer interaction is counted since the change of the VDW interaction energy within the strained layer is captured in the elastic energy of the substrate. The VDW energy difference calculated from the above method is 0.0019 eV per primitive unit cell of the MoS$_2$. As mentioned in the main text, this slip energy makes a negligible contribution to $\Delta E$. A similar order of magnitude of slip energy is expected in \{001\} epitaxy and other compliant substrate epitaxy systems.

\subsection{3. Detail strain analysis and minimization of $\Delta E$ in \{111\} and \{001\} compliant substrate epitaxy.}
As shown in Fig.~1(a) of the main text, the lattice mismatch between Au and MoS$_{2}$ is about 8\% biaxially in the case of \{111\} epitaxy. In this case, the Au film has a biaxial tensile strain ($\epsilon_{Au,11}$) in the growth plane and a normal strain component ($\epsilon_{Au,33}$) in the direction perpendicular to the plane. The shear strain is zero. The strain of the interface and surface, as defined in the main text, are same as the strain of the Au film in the growth plane. The strain for the MoS$_{2}$ is a biaxial compressive strain ($\epsilon_{MoS_{2},11}$) also with no shear component. With this form of the strain tensor given, the expression of $\Delta E$ can be written down based on Eqn.~(1) to (5) of the main text. Furthermore, the value of $\epsilon_{MoS_{2},11}$ is related to $\epsilon_{Au,11}$ through the condition of lattice matching at the interface:
\begin{equation}
a_{Au}(1 + \epsilon_{Au,11}) = a_{MoS_{2}}(1 + \epsilon_{MoS_{2},11})
\end{equation}
Once the number of Au layers is given, the only variables in the $\Delta E$'s expression are $\epsilon_{Au,11}$ and $\epsilon_{Au,33}$. The $\Delta E$ can then be minimized with respect to $\epsilon_{Au,11}$ and $\epsilon_{Au,33}$. Moreover, in this case, only the combined linear term ($f_{s}+f_{I}$), the combined quadratic terms ($C_{Sub,1111} + C_{Sub,1122} + C_{I,1111} + C_{I,1122}$) and the combined free interface and surface energies ($\gamma_{s}+\gamma_{I}$) enter into the minimized expression for $\Delta E$. (Due to the symmetry of the surface/interface, the linear term can be taken as scalar \cite{SurfaceStress}.) Thus the combined terms can be fitted as a whole to the DFT results.

In the case of \{001\} epitaxy, the lattice mismatch is about 8\% in direction \textbf{1} and -6\% in direction \textbf{2}, as shown in Fig.~1(b) of the main text. In the growth plane, the Au film has a pure tensile strain $\epsilon_{Au,11}^\prime$ in direction \textbf{1} and a pure compression strain $\epsilon_{Au,22}^\prime$ in direction \textbf{2}. The Au film also has a normal strain component $\epsilon_{Au,33}^\prime$ perpendicular to the growth plane. Similarly, the strain of the interface and surface are same as the strain in the Au film growth plane. In the MoS$_{2}$, likewise, it has normal compression strain $\epsilon_{MoS_{2},11}^\prime$ in direction \textbf{1} and normal tensile strain $\epsilon_{MoS_{2},22}^\prime$ in direction \textbf{2}. The shear components for both Au and MoS$_{2}$ remain zero. There are now two constraint equations relating the strain of Au and MoS$_{2}$ in the growth plane:
\begin{eqnarray}
a_{1(Au)}(1 + \epsilon_{Au,11}^\prime) = a_{1(MoS_{2})}(1 + \epsilon_{MoS_{2},11}^\prime)   \\
a_{2(Au)}(1 + \epsilon_{Au,22}^\prime) = a_{2(MoS_{2})}(1 + \epsilon_{MoS_{2},22}^\prime)
\end{eqnarray}
Once the number of Au layers is given, the variables in the total energy expressions are $\epsilon_{Au,11}^\prime$, $\epsilon_{Au,22}^\prime$ and $\epsilon_{Au,33}^\prime$. $\Delta E$ can then be minimized with respect to $\epsilon_{Au,11}^\prime$, $\epsilon_{Au,22}^\prime$ and $\epsilon_{Au,33}^\prime$. In this case, we have one combined linear term ($f_{s}+f_{I}$), two combined elastic constants ($C_{Sub,1111} + C_{I,1111}$ and $C_{Sub,1122} + C_{I,1122}$) and the combined free surface and interface energies ($\gamma_{s}+\gamma_{I}$) entering the minimized $\Delta E$ expression. Just as in the case of \{111\} epitaxy, these parameters are fitted as a whole to the DFT results.

In more detail, the fitting process includes both the energy and the in-plane strains through, These are incorporated within a sum of squares,  $\chi^2$:
\begin{equation}
\chi^2 = \emph{w} ~ \sum_{i}(\Delta E[i] - \Delta \tilde{E}[i])^2 + (1-\emph{w}) ~ \sum_{j}(\epsilon[j] - \tilde{\epsilon}[j])^2.
\end{equation}
In the above equation, $\Delta E[i]$ is the DFT calculated excess energy of Au-MoS$_{2}$ system with $i$ layers of Au (defined by Eqn.~(6) of the main text). $\Delta \tilde{E}[i]$ is the excess energy from the elasticity model also with $i$ layers of Au (the minimized Eqn.~(1) of the main text). The strain sum term is defined similarly. For \{111\} epitaxy, the in-plane biaxial strain comes into the strain sum. For \{001\} epitaxy, both $\epsilon_{11}$ and $\epsilon_{22}$ are included the strain sum. The expressions for the $\tilde{\epsilon}[j]$ are also obtained from DFT results. The indices $i$ and $j$ run over all the computed DFT data. $\emph{w}$ is a weighting parameter that can be used to tune the relative importance of the fit to the energy versus the fit to the strain. This parameter is arbitrary. Since the DFT calculated energy is more accurate the DFT calculated strain, we choose to put more weight ($w=2/3$) on the energy. The fitting process minimizes $\chi^2$ with respect to these unknown parameters in the elasticity theory model. The fitted results for both orientations are summarized in Table~\ref{tab:table2}.
\begin{table}[h]
\caption{\label{tab:table2}%
The values of fitting parameters \footnote{All fitting parameters are in unit of [eV/\AA$^2$]}.
}
\begin{ruledtabular}
\begin{tabular}{ccccc}
\textrm{ \hspace{20 mm} } & \textrm{Linear terms} & \multicolumn{2}{c}{\textrm{Quadratic terms}}                                      & \textrm{Free energy terms} \\ \hline
\multirow{2}{*}{ \{111\} epitaxy} & $f_{s}+f_{I}$ & \multicolumn{2}{c}{$C_{Sub,1111} + C_{Sub,1122} + C_{I,1111} + C_{I,1122}$} & $\gamma_{i}+\gamma_{s}$ \\
                                & 0.167 & \multicolumn{2}{c}{3.988}                                                   & 0.067 \\ \hline
\multirow{2}{*}{ \{001\} epitaxy} & $f_{s}+f_{I}$ & $C_{Sub,1111} + C_{I,1111}$ & $C_{Sub,1122} + C_{I,1122}$ & $\gamma_{i}+\gamma_{s}$ \\
                               &  -0.541   & 45.703                        & 45.656                       & 0.099
\end{tabular}
\end{ruledtabular}
\end{table}

\begin{figure}[h]
 \includegraphics{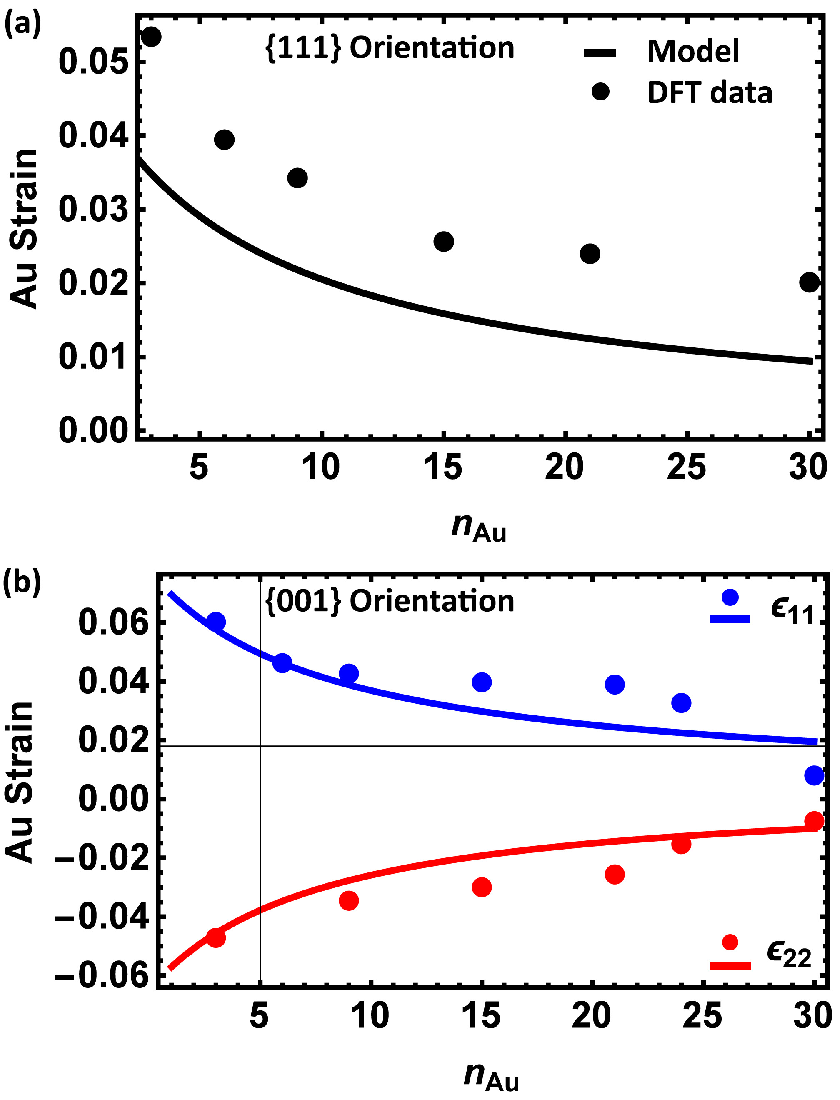}%
  \caption{\label{fig:fig3} (color online) The fitted Au strain curve and DFT Au strain data for (a) the biaxial strain in \{111\} epitaxy and (b) the two in-plane strains in \{001\} epitaxy.}
\end{figure}

The fitted energy curves are shown in the Fig.~4 of the main text. The fitted strain curves are shown in Fig.~\ref{fig:fig3}. The strain curves are also in reasonable agreement with the DFT predictions.

\bibliography{Zhousupplemental}